\documentclass[a4paper,11pt]{article}
\usepackage{amsmath,amssymb,amsfonts}
\usepackage{booktabs}
\usepackage{graphicx}
\usepackage{subfigure}
\usepackage{color}

\newtheorem{remark}{Remark}

\newcommand{\revised}[1]{\textcolor{black}{#1}}
\renewcommand{\aa}{r}
\newcommand{\R}{\mathbb{R}}
\newcommand{\fer}[1]{(\ref{#1})}
\newcommand{\ve}{\varepsilon}
\newcommand{\e}{\delta}
\newcommand{\be}{\begin{equation}}
\newcommand{\ee}{\end{equation}}
\newenvironment{equations}{\equation\aligned}{\endaligned\endequation}
\begin{document}
\title{Kinetic models for optimal control of wealth inequalities}
\author{Bertram D{\"u}ring\thanks{Email:~bd80@sussex.ac.uk, Department of Mathematics, University of Sussex, Pevensey~II, Brighton, BN1 9QH, United Kingdom}
\and Lorenzo Pareschi\thanks{Email:~lorenzo.pareschi@unife.it, Dipartimento di Matematica e Informatica
Via Machiavelli 35, 44121 Ferrara, Italy}
\and Giuseppe Toscani\thanks{Email:~giuseppe.toscani@unipv.it, Dipartimento di Matematica and IMATI, CNR,
Via Ferrata 1, 27100 Pavia, Italy}} 
\maketitle
\begin{abstract}
\noindent We introduce and discuss optimal control strategies for kinetic models for wealth distribution in a simple market economy, acting to minimize the variance of the wealth density among the population. Our analysis is based on a \revised{finite time horizon approximation, or model predictive control, of the corresponding control problem for the microscopic agents' dynamic and results in an alternative theoretical approach to the taxation and redistribution policy at a global level}.  It is shown that in general the control is able to modify the Pareto index of the stationary solution of the corresponding Boltzmann kinetic equation, and that this modification
can be exactly quantified. Connections between previous Fokker-Planck
based models and taxation-redistribution policies and the present approach are also discussed.

\end{abstract}

\noindent{\bf Keywords}:
Wealth distribution, wealth inequalities, kinetic models, optimal control, finite time
horizon, Fokker-Planck equations, Pareto tails, taxation,
redistribution.
% %% keywords here, in the form: keyword \sep keyword
% %% MSC codes here, in the form: \MSC code \sep code
% %% or \MSC[2008] code \sep code (2000 is the default)

%\tableofcontents

\section{Introduction}

Any society with a growing reliance on capital experiences an increasing concentration of wealth, which leads in general to a marked social inequality.  How to reduce these social inequalities in capitalistic countries is a debated issue.  The usual government policies are to use proportional taxation, with the expectation  that  a  progressive tax
system would prevent excess concentration of wealth.  A recent approach to this relevant economic question can be found in  Piketty \cite{Piketty}, whose main conviction is that the effect of the tax on capital income is not only to reduce
the total accumulation of wealth, but to modify the structure of the wealth distribution over the long run.
In other words,  a confiscatory tax on high incomes combined with a progressive tax on the value of the capital is viewed by Piketty as the only way to prevent the natural tendency of capitalism to head towards excessive inequality. 

As a matter of fact, long term predictions on economic systems are very difficult to justify, and a serious debate would require a rigorous analysis based on well-established models of wealth distribution. In this developing area of research, mathematical modeling of economic systems has had interesting advances in recent years \cite{DJR, DuMaTo08, DMT09, ParTosBook}.

Starting from the pioneering studies of Angle \cite{Angle}, most of
these models  sink their roots into statistical mechanics \cite{Yakovenko,Donadio}, and are based on methods borrowed from the kinetic theory of rarefied gases and the Boltzmann equation \cite{Cerc88, CIP94}. The main original motivation at the basis of this modeling was  to understand the possible reasons of formation of heavy tails in the distribution of wealth, as predicted by the economic analysis of the Italian economist Vilfredo Pareto \cite{Par}.  

One of the kinetic models of wealth distribution able to reproduce the formation of Pareto tails on the basis of few physically plausible hypotheses has been introduced in 2005 in \cite{CPT05}. There, the evolution of wealth has been based on binary trades modeled to include the idea that wealth changes hands for a specific reason: one agent intends to invest their wealth in some asset, property etc.\ in possession of their trade partner. Typically, such investments bear some risk, and either provide the buyer with some additional wealth, or lead to the loss of wealth in a non-deterministic way. An easy realisation of this idea consists in coupling the saving propensity parameter \cite{Cha,CC} with some risky investment that yields an immediate gain or loss proportional to the current wealth of the investing agent.
Leaving the details of the microscopic trade to Section~\ref{ist}, we recall here that the model for wealth distribution introduced in \cite{CPT05} revealed to be very flexible with respect to the addition of further economic aspects, including the possibility of studying the effects of taxation and redistribution \cite{Bisi, BST, To09},  the role and consequences of the addition of a parameter describing agent's knowledge \cite{PT-kno},  and the possibility to use the kinetic interaction operator to construct suitable equations of hydrodynamics \cite{DuTo07, To17}. 

Going back to the problem of capitalistic societies and wealth
inequality, it is interesting to remark that the numerical simulation
of the evolution of the kinetic model for wealth and knowledge
developed in  \cite{PT-kno},  led to the conclusion that the unequal
distribution of knowledge in a multi-agent society is itself a cause
of an unequal distribution of wealth among agents.  Other aspects of
wealth inequality and surplus theory have been recently analysed from
the mathematical point of view \cite{PR}, with the aim to  to find a
relationship between agents' risk aversion and inequality of incomes. These studies clearly outline the importance of resorting to mathematical modeling to test and eventually verify economical hypotheses.

In this paper, we will discuss a possible alternative to the standard taxation and redistribution rules, which relies on a suitable control applied to the microscopic trades describing the wealth distribution of the multi-agent system. Recent applications of control problems to kinetic models with binary interactions describing opinion formation can be found in \cite{AHP, APZ} (cf. also \cite{AlbiEtAl16} for an exhaustive review). Indeed, the possibility  to effectively exercise a control on opinion and to evaluate the impact of modern communication systems,
like social networks, to the dynamics of opinions, is a challenging problem of increasing importance. 
%The main ideas developed in  \cite{AHP, APZ} can be naturally adapted to the present situation concerning wealth distribution.  

\revised{More precisely, we assume the existence of a policy maker (a government or a local administrator) that applies a suitable control process to each economic interaction with the aim to minimize a given cost functional measuring the wealth inequalities in the system. This control acts as an agent dependent taxation/redistribution dynamic and, for the sake of simplicity, it is assumed conservative over the whole set of agents so that the total amount of wealth remains unchanged. The resulting constrained dynamic takes the form of an optimal control problem which, for a large set of agents, turns out to be computationally prohibitive due to its intrinsic complexity and therefore approximate solution are sought even if suboptimal. Among various possible approaches here, following \cite{AHP, APZ}, we apply a finite time horizon strategy based on model predictive control. In the simpler case of instantaneous control the problem can be solved explicitly giving rise to a feedback control that can be embedded in the microscopic system.}

\revised{By considering binary interactions, the application of this feedback control can be
shown to change the saving propensities of the agents, which induces a
smaller variance for the density of wealth of the population. For the binary dynamic  introduced in \cite{CPT05} the corresponding feedback control originates a Boltzmann equation whose
stationary states, compared to the original uncontrolled model,  have a larger Pareto index. An
explicit result in the direction of Piketty's opinion 
\cite{Piketty} is that, in the \emph{quasi-invariant interaction
  limit}, among others, we can recover the same Fokker--Planck equation
resulting from a standard taxation and redistribution process \cite{Bisi, BST}.} 

The rest of the manuscript is organized as follows. In Section 2 we introduce the microscopic model in the optimal control setting. For this model we derive the explicit feedback control in a finite time horizon approximation and focus on the binary interaction case. Section~3 is devoted to the study of the corresponding kinetic models. We focus on the CPT model \cite{CPT05} and show that the action of the control is capable to increase the Pareto index of the corresponding wealth distribution, thereby reducing inequalities. To have a further insight in the stationary states of the system, in Section 4 we pass to the limit controlled Fokker-Planck equation and show how it can be reinterpreted as a taxation-redistribution model. Some numerical simulations which confirm our analysis are also reported. 

\section{Optimal control of wealth inequalities}

\subsection{A microscopic model with control}
Let us consider the microscopic evolution of the wealths of $N$ agents, where each agent's wealth
$w_i$,  $i = 1, . . . , N,$ evolves according to the following first order dynamical system
\begin{align}
\dot w_i(t) &= \frac1N \sum_{j=1}^N a_{ij} (w_j-w_i)+ u_i, \qquad w_i (t=0) = w_{i,0}\ge 0.
\label{eq:1}
\end{align}
In \fer{eq:1} the nonnegative constants $a_{ij}$ define the exchange parameters of the trades and the $u_i$'s are control terms. In general, to ensure the positivity in time of the wealth variables, it is assumed that the exchange parameters satisfy $a_{ij} <1$ for $i,j =1,2,\dots, N$.

The controls $u_i$ act in order to redistribute wealth with the aim to
decrease the variance of wealth among agents. \revised{This can be achieved by
minimizing the functional  
\be
\arg\min_{u\in\cal U} J(w,u)=\frac12 \int_0^T \frac1{N} \sum_{j=1}^N \left(L_j(w)+ \nu|u_j|^2\right)\,dt,
\label{eq:1b}
\ee
where  $\cal U$  is the space of admissible controls,
$w=(w_1,\ldots,w_N)$, $u=(u_1,\ldots,u_N)$ and $L_j(w)$ is a target cost functional which measures the level of wealth inequalities in the system.}

\revised{An example is given by
\be
L_j(w)=\frac1{N} \sum_{k=1}^N |w_j-w_k|^m,\quad m \geq 1,
\label{eq:cost}
\ee
where for $m=2$ we have a classical quadratic cost functional which corresponds to minimize the variance of the wealth among all agents.}

The constant $\nu  > 0$ is
a selective penalization parameter which takes into account that we
may want to apply different taxation rules to different level of
incomes. As we shall discuss later on, since the control essentially
acts on interactions among agents of the system, the constant $\nu$
can be assumed to depend on the frequency of exchanges. In this way,
the control $u$ can be understood as the external action of a
government which aims to reduce inequalities, by acting on exchanges, through wealth-dependent taxation and redistribution among agents.

\revised{Problem \eqref{eq:1}-\eqref{eq:1b} can be reformulated as Mayer's problem and solved by dynamic programming or Pontryagin's maximum principle \cite{krstic1995, Sontag1998aa}. However, the main drawback relies on the fact that the equation for the adjoint variable has to be solved backwards in time  over the full time interval $[0,T]$. In particular,  for large values of $N$ the computational effort becomes prohibitive. Also, assuming $u=\mathcal{A}(x)$ where $\mathcal{A}$ fulfills a Riccati differential equation cannot be pursued here due to the  large dimension of $\mathcal{A} \in \mathbb{R}^{N \times N}$ and a possible general nonlinearity in the coefficients $a_{ij}$ (see \cite{HSP}).  A standard methodology, when dealing with such complex system, is based on model predictive control where instead of solving the control problem over the whole time horizon, the system is approximated by an iterative solution over a sequence of finite time steps \cite{CaBo:04}. }

\subsection{Instantaneous control} \label{ist}
%Due to its intrinsic difficulties for large values of $N$, the above problem is tackled in a finite time %horizon approximation. 
\revised{We derive a feedback control $u$ based on a finite time horizon strategy. This feedback control will in general only be suboptimal. Rigorous results on  the properties of $u$ for quadratic cost functional and linear and nonlinear dynamics  are available, for example, in \cite{CaBo:04}. The receding horizon framework applied here is also called instantaneous control in the engineering literature.} 

Following the approach in \cite{AlbiEtAl16}, we assume a finite time horizon $\Delta t \le 1$ and in a time-discrete setting with times $t^n=n\Delta t$ we consider the problem
\begin{align}
w_i^{n+1} &= w_i^{n} +\frac{\Delta t}N \sum_{j=1}^N a_{ij}
            (w_j^n-w_i^n)+\Delta t \,u_i^n.
            \label{eq:2}
\end{align}
\revised{In this case we are led to minimize the cost functional
\be
J_{\Delta t}(w,u)=\frac 1{2N}\sum_{j=1}^N\left( 
L_j(w^{n+1})+ {\nu} |u_j^n|^2\right).
\ee
Let us first consider the case of a quadratic cost functional, namely \eqref{eq:cost} in the case $m=2$.}

The necessary optimality conditions (which can be obtained by direct differentiation with respect to $u_i^n$) yield
$$
\frac {\Delta t}{N^2}\sum_{j,k=1}^N
\bigl(w_j^{n+1}-w_k^{n+1}\bigr)\bigl(
\delta_{ij} - \delta_{ik}\bigr)+\frac\nu{N}  u_i^n=0,
$$ 
where as usual $\delta_{ij}$ denotes the Kronecker delta.

Solving for the controls $u_i^n$ we get
\be
u_i^n=-\frac{2\Delta t}{\nu N}\sum_{j=1}^N
\bigl(w_i^{n+1}-w_j^{n+1}\bigr)=-\frac{2 \Delta t}{\nu }
\bigl(w_i^{n+1}-\bar w^{n+1}\bigr),
\ee
where $\bar w^{n+1} = \sum_{j=1}^Nw_j^{n+1}/N $ denotes the mean wealth of the agents at time $(n+1) \Delta t$. Note that the above controls satisfy the identity
$$
\sum_{i=1}^N  u_i ^n = -\frac{2\Delta t}{\nu }
\sum_{i=1}^N \bigl(w_i^{n+1}-\bar w^{n+1}\bigr)=0,
$$
which implies that all taxes are redistributed among agents.

Using the discrete dynamics (\ref{eq:2}) we finally obtain the explicit expressions
\begin{align}
u_i^n=&-\frac{2\Delta t}{\nu + 2\Delta t^2}\left(
w_i^{n}-\bar w^{n} +\frac{\Delta t}N \sum_{j=1}^N a_{ij}
            (w_j^n-w_i^n)-\frac{\Delta t}{N^2} \sum_{j,k=1}^N a_{kj}(w_j^n-w_k^n)\right).
            \label{eq:3}
\end{align}
Expression (\ref{eq:3}) furnishes a feedback control for the fully discretized problem, which
can be plugged as an instantaneous control into (\ref{eq:2}). Note, however, that the instantaneous control (\ref{eq:3}) in the discretized dynamics (\ref{eq:2}) is of order
$\mathcal{O}(\Delta t)$. In order to obtain an effective contribution of the control in the dynamics
we will make some further natural assumptions. First, we assume that the penalization parameter $\nu$ scales with the time discretization as $\nu =2\gamma \Delta t$. This is consistent with the idea that for very short time horizons we need a stronger control to achieve the desired goal. Second, if one agrees with the fact that a control on wealth has to depend also on the frequency and intensity of interactions, one is lead to assume that the parameter $\gamma$ has to depend  on the sum $A$ of the exchange parameters $a_{ij}$, and it is inversely proportional to $A$. This guarantees that, in absence of exchanges in the system, the control on wealth looses its meaning.   

In this way the instantaneous controls reads
\begin{align}
u_i^n=&-\frac{1}{\gamma +\Delta t}\left(
w_i^{n}-\bar w^{n} +\frac{\Delta t}N \sum_{j=1}^N a_{ij}
            (w_j^n-w_i^n)-\frac{\Delta t}{N^2} \sum_{j,k=1}^N a_{kj}(w_j^n-w_k^n)\right).
            \label{eq:4}
\end{align}

In the above setting, if we assume $a_{ij}=a_{ji}$, the mean wealth is conserved, so that $\bar w^{n+1}= \bar w$, and the minimization of the functional $J_{\Delta t}(w,u)$ corresponds to minimize the quadratic inequality indicator
\be
G_2=\frac{\sum_{j,k=1}^N (w_j-w_k)^2}{2N^2{\bar w}^2}.
\ee
Note that a standard indicator of wealth inequality, closely related
to the one above, is the Gini coefficient, defined as
\be
G_1=\frac{\sum_{j,k=1}^N |w_j-w_k|}{2N^2{\bar w}}.
\label{eq:Gini}
\ee
\revised{In our setting, minimization of the Gini coefficient corresponds to the cost functional \eqref{eq:cost} for $m=1$.}
Analogous computations show that this choice leads to the feedback control
\be
u_i^n=-\frac{2\Delta t}{\nu N}\sum_{j=1}^N
\frac{(w_i^{n+1}-w_j^{n+1})}{|w_i^{n+1}-w_j^{n+1}|},
\ee
where again we have $\sum_{i} u_i^n =0$ and therefore all taxes are redistributed. In this case, however, even using the expression of the dynamic \eqref{eq:2} it is not possible to give an explicit expression to the above control term. \revised{Similar conclusions are obtained for $m > 2$.}
% In the sequel we will focus  where we have the feedback control
%\be
%u_i^n=-\frac{m\Delta t}{\nu N}\sum_{j=1}^N
%|w_i^{n+1}-w_j^{n+1}|^{m-2}\bigl(w_i^{n+1}-w_j^{n+1}\bigr).
%\ee
%}

\begin{remark} A more realistic dynamic typically includes a random part into the evolution of the wealth system, which now reads
\begin{align}
\dot w_i(t) &= \frac1N \sum_{j=1}^N a_{ij} (w_j-w_i)+ \eta_iw_i + u_i, \qquad w_i (t=0) = w_{i,0}\ge 0.
\label{eq:risk}
\end{align}
In \fer{eq:risk}, the $\eta_i$, $i=1,2,\dots, n,$ denote a sequence of
independent and identically distributed random variables such that
$\langle \eta_i\rangle =0$ and  $\langle \eta_i^2\rangle =\sigma$,
where $\langle \cdot\rangle$ denotes mathematical expectation. The
additional random part represents risks which are always present in economic trades \cite{ParTosBook}. It is reasonable, however, to assume that the control could act only on the deterministic part of the evolution. %\revised{For example, in the case $p=2$, giving again the expression \fer{eq:4}.} 
\end{remark}

\subsection{Control of binary interactions}\label{bin}

The special case $N=2$ describes binary interactions. Binary interactions are at the basis of the kinetic description of wealth distribution in multi-agent systems \cite{ParTosBook}. In absence of risky components we obtain
\begin{align}
\nonumber
w_i^{n+1} &= w_i^{n} + {\Delta t}\, \bar a_{ij}(w_j^n-w_i^n)+\Delta t \,u(w_i^n,w_j^n),\\[-.2cm]
\label{eq:2b}
\\[-.2cm]
\nonumber
w_j^{n+1} &= w_j^{n} + {\Delta t}\, \bar a_{ji}(w_i^n-w_j^n)+\Delta t \,u(w_j^n,w_i^n),
\end{align}
where $\bar  a_{ij} =  a_{ij}/2$ for every $i\not= j$. \\
\revised{
In the case of a quadratic cost functional we have 
\begin{align}
u(w_i^n,w_j^n)=&-\frac{1}{\gamma +\Delta t}\biggl(
\frac12(w_i^n-w_j^n) +\frac{\Delta t}2 (\bar a_{ij}+\bar a_{ji}) (w_j^n-w_i^n)\biggr).
            \label{eq:4b}
\end{align}}
Note that in the above formulation both the dynamics as well as the control functional operate at the level of the binary interaction pair $(w_i,w_j)$. Note again that the binary dynamics preserves the local mean wealth if and only if $\bar a_{ij}= \bar a_{ji}$. 
If we now define $p=\Delta t\, \bar a_{ij}$, $q= \Delta t\, \bar a_{ji}$ we can write the controlled binary Boltzmann dynamics for the pair $(v,w)$ in the form 
\begin{equations}\label{ori}
  v^*&= v + p (w-v)+\Delta t \, u(v,w),\\
w^*&= w + q(v-w)+\Delta t\, u(w,v),
\end{equations}
with
\begin{align}
 u(v,w)=& \frac{1}{2(\gamma + \Delta t)}(1-p-q)(w-v).
            \label{eq:5b}
\end{align}
Collecting all terms together, and setting
 \be
\beta =  \frac{\Delta t}{\gamma + \Delta t},
\label{eq:beta}
 \ee
 the binary relations \fer{ori} can be rewritten as
\begin{equations}\label{co2}
  v^*&=v + \Bigl(p+\frac{\beta}{2}(1-p-q)\Bigr)(w-v) 
      =v + \tilde p(w-v),\\
w^*&=w + \Bigl(q+\frac{\beta}{2}(1-p-q)\Bigr)(v-w) 
    =w + \tilde q( v-w).
\end{equations}
Hence, we observe that in the binary case the feedback control can be reformulated as a modification of the original  mixing coefficients of the binary interaction. Note that, since by definition both $p$ and $q$ are less than one, and $0 < \beta < 1$ \revised{(where $\beta=0$ coincides with absence of control and $\beta=1$ yields  maximum control)}, the new mixing coefficients $\tilde p$ and $\tilde q$ still satisfy $0 < \tilde p, \tilde q <1$. 

\revised{
It is interesting to consider the model predictive control approximation originated by the minimization of the cost functional for $m=1$ in the case of binary interactions. In this case, in fact, assuming $\nu=2\gamma\Delta t$ we have the implicit control definition
\be
u(w_i^n, w_j^n) = -\frac{1}{2\gamma} \frac{(w_i^{n+1}-w_j^{n+1})}{|w_i^{n+1}-w_j^{n+1}|}.
\ee 
Now setting $z_{ij}^n=w_i^n-w_j^n$ from the binary interaction dynamic \eqref{eq:2b} we obtain the nonlinear equation
$$
z_{ij}^{n+1} = z_{ij}^n (1-\Delta t(\bar{a}_{ij}-\bar{a}_{ji}))-\Delta t \frac{z_{ij}^{n+1}}{\gamma|z_{ij}^{n+1}|}.
$$ 
It is easy to verify that the above equation admits a solution only for 
$$
 |z_{ij}^n| \geq \frac{\Delta t}{\gamma(1-\Delta t(\bar{a}_{ij}-\bar{a}_{ji}))}.
$$
Now using the same notations as in \eqref{ori} we have the explicit feedback control
\be
u(v,w) = \left\{
\begin{array}{cc}
 \displaystyle-\frac{1}{2\gamma}, & \displaystyle v \geq w +\frac{\Delta t}{\gamma(1-(p-q))},  \\[+.3cm]
 \displaystyle\frac{1}{2\gamma}, & \displaystyle v \leq w -\frac{\Delta t}{\gamma(1-(p-q))},  \\[+.3cm]
 0, & {\rm otherwise}.   \\
\end{array}
\right.
\ee
Therefore, a fixed taxation amount is applied to the richer (and redistributed to the poorer) of the two agents only if the difference in wealth is above a certain threshold. Note that, the taxation process is such that $v^*\geq 0$ and $w^* \geq 0$ and that the resulting dynamic cannot be reformulated as a modification of the original mixing coefficients of the binary interaction as in \eqref{co2}. 
}

\section{Boltzmann models for wealth distribution with control}

The basic model discussed in this section has been introduced 
in 2005 in \cite{CPT05} within the framework of classical models of wealth distribution in economy, to understand the possible formation of 
heavy tails, as predicted by the economic analysis of the Italian economist Vilfredo Pareto \cite{Par}.
This model belongs to a class
of models in which the interacting agents are indistinguishable. In most of these
models an agent's \emph{state} at any instant of
time $t\geq0$ is completely characterized by his current wealth
$v\geq0$ \cite{DuMaTo08,DMT09}.  When two agents encounter in a trade, their {\em pre-trade
wealths\/} $v$, $w$ change into the {\em post-trade wealths\/}
$v^*$, $w^*$ according to the rule \cite{Cha,CC}
\[
%  \label{eq.trules}
  v^* = p_1 v + q_1 w, \quad w^* = q_2 v + p_2 w.
\]
The {\em interaction coefficients\/} $p_i$ and $q_i$ are
non-negative random variables. While $q_1$ denotes the fraction of
the second agent's wealth transferred to the first agent, the
difference $p_1-q_2$ is the relative gain (or loss) of wealth of the
first agent due to market risks. It is usually assumed that $p_i$
and $q_i$ have fixed laws, which are independent of $v$ and $w$, and
of time. This means that the amount of wealth an agent contributes
to a trade is (on the average) proportional to the respective
agent's wealth.

\subsection{The control of the Cordier-Pareschi-Toscani (CPT) model} \label{CPT}

In \cite{CPT05}  the trade has been modelled to include the idea that wealth changes hands for a specific reason: one agent intends to {\em invest\/}\nobreakspace his wealth in some asset,
property etc.\ in possession of his trade partner. Typically, such investments bear
some risk, and either provide the buyer with some additional wealth, or lead to the
loss of wealth in a non-deterministic way. An easy realization of this idea consists in coupling a constant saving propensity parameter \cite{Cha,CC} with some {\em risky
investment\/} that yields an immediate gain or loss proportional to the current wealth
of the investing agent
\begin{equations}
  \label{eq.cpt}
  v^* &= v + \frac{1-\lambda}2(w-v) +\eta_1 v , \\
  w^* &= w + \frac{1-\lambda}2(v-w) +\eta_2 w,
\end{equations}
where $0 <\lambda <1$ is the parameter which identifies the saving propensity, namely the intuitive behavior which prevents the agent to put in a single trade  the whole amount of his money. In this case
 \[
 \label{cpt1}
   p_i =  \frac{1+\lambda}2 + \eta_i , \quad q_i = \frac{1-\lambda}2 \quad (i=1,2).
  \]
As specified above, the coefficients $\eta_1,\eta_2$ are random parameters, which are
independent of $v$ and $w$, and distributed so that always
$v^*,\,w^*\geq  0$, i.e.\ $\eta_1,\,\eta_2\geq -(1+\lambda)/2$.

\begin{figure}
\centering
\includegraphics[width=0.6\textwidth]{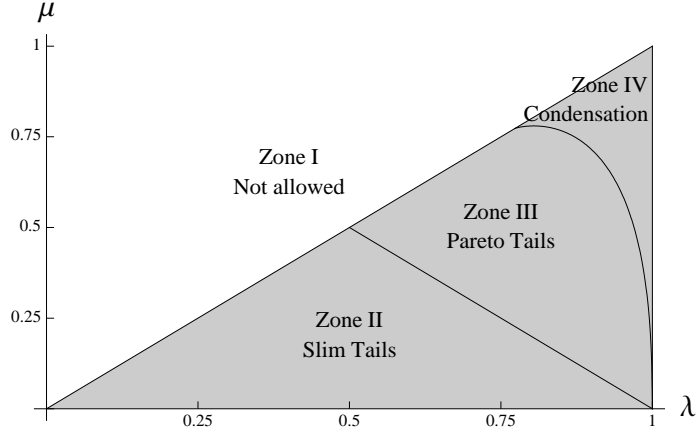}
\caption{Parameter ranges with different tail regimes in the CPT model
  in the $\lambda$-$\mu$-plane.}
\label{fig:cptzones}
\end{figure}

Owing to classical arguments of kinetic theory \cite{ParTosBook},  it has been shown in \cite{CPT05} that the evolution of the wealth density consequent to the binary interactions \fer{eq.cpt} obeys a Boltzmann-type equation.  Let us denote with $f(v,t)$   the distribution of the agents wealth $v \ge 0$ at time $t >0$. Then, the equation for the evolution of $f(v,t)$ can be fruitfully written in weak form.
It corresponds to say that, for any smooth function $\phi$, $f$ satisfies the equation
\begin{multline}
\frac{d}{dt} \int_{\R_+}\phi(v) f(v,t) dv= \\
\frac{1}{2}\left\langle \int_{\R_+ \times\R_+} f(v,t)f(w,t) \big( \phi(v^*)+\phi(w^*)-\phi(v)-\phi(w)\big)\, dv dw \right\rangle. \label{bilcinetica}
\end{multline}  
A simple computation shows that, unless the
random variables are centered, i.e.\
$\langle\eta_1\rangle=\langle\eta_2\rangle=0$,  the mean wealth is not preserved, but it increases or
decreases exponentially (see the computations in \cite{CPT05}).
For centered $\eta_i$,
\[
%\label{con}
  \langle v^* + w^* \rangle = (1+\langle\eta_1\rangle) v
  + (1+\langle\eta_2\rangle) w = v + w ,
\]
implying conservation of the average wealth, so that
 \[
 m(f) = \int_{\R^+}v f(v,t) \, dv = m(f_0).
 \]
Various specific
choices for the $\eta_i$ have been discussed in \cite{MaTo08}. The
easiest one leading to interesting results is $\eta_i=\pm \mu$, where
each sign comes with probability $1/2$. The factor
$\mu\in(0,\lambda)$ should be understood as the {\em intrinsic
risk\/} of the market: it quantifies the fraction of wealth agents
are willing to gamble on. Within this choice, one can display the
various regimes for the steady state of wealth in dependence of
$\lambda$ and $\mu$, which follow from numerical evaluation. In the
zone corresponding to low market risk, the wealth distribution shows
again \emph{socialistic} behavior with slim tails. Increasing the
risk, one falls into \emph{capitalistic}, where the wealth
distribution displays the desired Pareto tail. A minimum of saving
($\lambda>1/2$) is necessary for this passage; this is expected since
if wealth is spent too quickly after earning, agents cannot
accumulate enough to become rich. Inside the capitalistic zone, the
Pareto index decreases from $+\infty$ at the border with
\emph{socialist} zone  to unity. Finally, one can obtain a steady
wealth distribution which is a Dirac delta located at zero. Both
risk and saving propensity are so high that a marginal number of
individuals manages to monopolize all of the society's wealth. In
the long-time limit, these few agents become infinitely rich,
leaving all other agents truly pauper. One
obtains four zones as depicted in Figure~\ref{fig:cptzones}. Note that Zone 1 is not allowed since $|\mu|<\lambda$.

Using the notations of Section \ref{bin} we can solve the control problem for the CPT-model with risk 
\begin{equations}\label{ori2}
  v^{**}&=v + p(w-v) +  \Delta t\, u(v,w) +  \Delta t\, \eta_1 v,\\
w^{**}&= w + q(v-w)+ \Delta t\, u(w,v)+ \Delta t\, \eta_2 w,
\end{equations}
where 
 \be\label{val}
 p= q =  {\Delta t}\, \frac{1-\lambda}2 = \alpha (1-\lambda),\qquad \alpha=\frac{\Delta t}2. 
 \ee
\revised{In the case of a quadratic cost functional we obtain as} feedback control on the deterministic part the value
 \be\label{conCPT}
 u(v,w) =\frac {\beta}{4\alpha}\, \left( 1- 2\alpha (1-\lambda) \right)(w-v).
 \ee
Consequently the post-control interaction \fer{co2} has deterministic interaction coefficients
 \be\label{post}
 \tilde p = \alpha ({1-\lambda}) + \frac{\beta}{2}\left( 1- 2\alpha(1-\lambda)\right) = \frac{\beta}{2}+\alpha(1-\lambda)(1-\beta).
 \ee
Finally, if we assume $\Delta t =1$ ($\alpha=1/2$), we can write the controlled binary interactions
\begin{equations}\label{CPTc}
v^{**}&=v + \left(\frac{1-\lambda(1-\beta)}2\right)(w-v) + \eta_1 v,\\
w^{**}&= w + \left(\frac{1-\lambda(1-\beta)}2\right)(v-w)+ \eta_2 w,
\end{equations}
where now $\beta=1/(1+\gamma)$. Note that negative values of the wealth are now avoided if $\mu \in (0,\lambda(1-\beta))$ for $\lambda (1-\beta) > 1/2$. This gives an upper bound for the maximum admissible control $\beta < 1-1/(2\lambda)$.

\begin{figure}[t]
\centering
\includegraphics[width=0.48\textwidth]{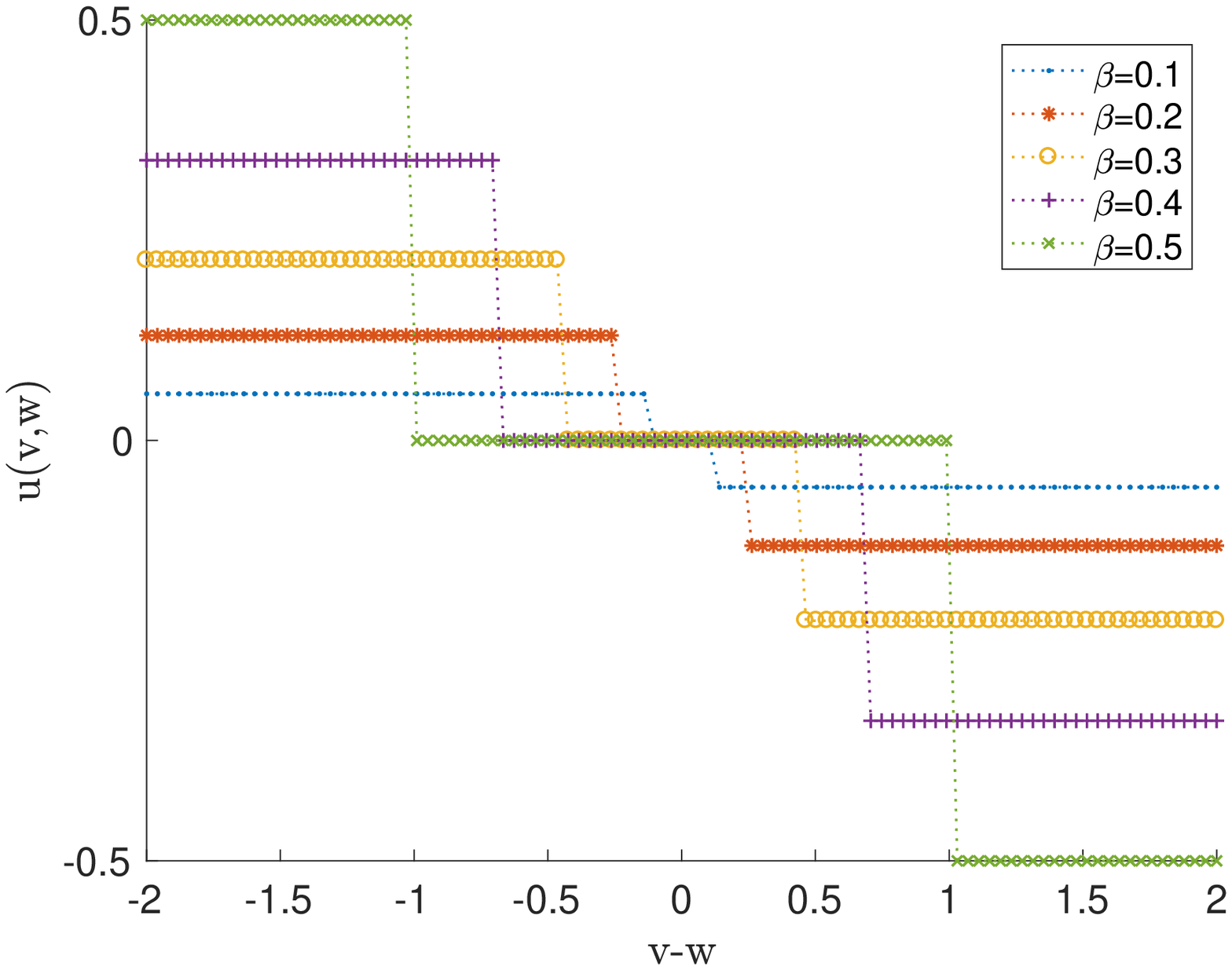}\,\,%
\includegraphics[width=0.48\textwidth]{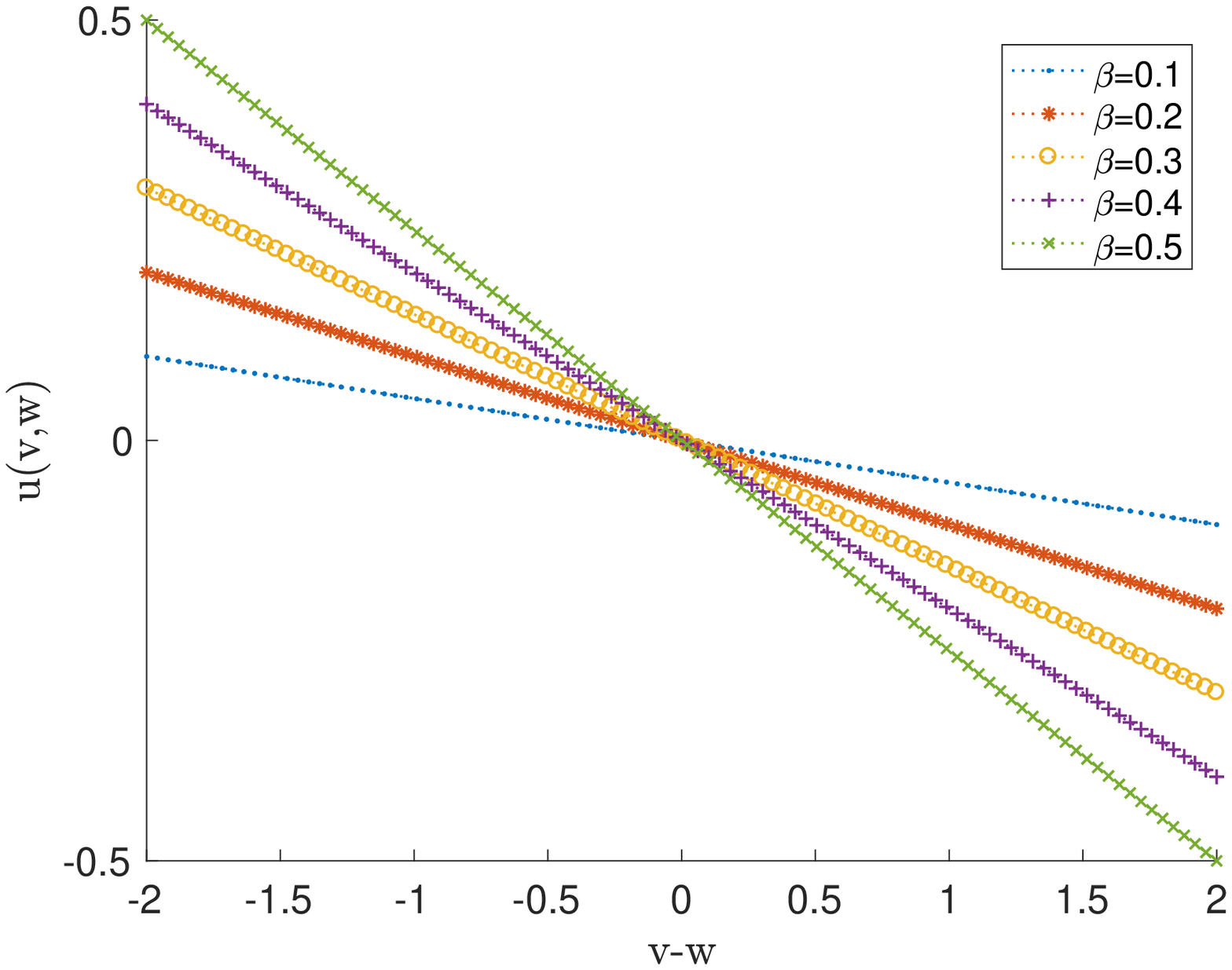}
\caption{The different feedback controls for $m=1$ (left) and $m=2$ (right) for various values of the penalization parameter $\beta$. Here $\alpha=1/2$ and $\lambda=1$.}
\label{fig:control}
\end{figure}

\revised{In a similar way, if we consider the explicit control obtained for the cost functional \eqref{eq:cost} for $m=1$ using \eqref{eq:beta} we have for deterministic part of the CPT model
\be
u(v,w) = \left\{
\begin{array}{cc}
-\displaystyle\frac{\beta}{4\alpha(1-\beta)}, & \displaystyle v \geq w +\frac{\beta}{(1-\beta)},  \\[+.3cm]
 \displaystyle\frac{\beta}{4\alpha(1-\beta)}, & \displaystyle v \leq w -\frac{\beta}{(1-\beta)},  \\[+.3cm]
 0, & {\rm otherwise}.   \\
\end{array}
\right.
\label{conCPT2}
\ee
In presence of noise positivity of the wealth is guaranteed for $\mu \in (0,\lambda/2)$ and all values of $\beta < 1$ are admissible. It should be noted, however, that large values of $\beta$ implies a stronger control but over a smaller number of agents (see Figure \ref{fig:control}, left).  
}

%\be
%\tilde p = \frac{1-\lambda}{2}+\frac{\lambda\beta}{2} = p +\frac{\lambda\beta}{2}.
%\ee 
 
%Finally, if we assume that the penalisation parameter $\gamma$ is inversely proportional to the intensity of the mixing parameters
% \be\label{ipo1}
% \gamma = \frac \vartheta{1-\lambda},
% \ee
% we obtain the expression
%  \be\label{post1}
% \tilde p = \tilde q =  \alpha ({1-\lambda}) \left( 1 + \frac{1- 2\alpha(1-\lambda)}{\vartheta +2\alpha(1-\lambda)}\right).
% \ee

%%%
\subsection{Control and Pareto tails}

The formation of stationary states and their properties have been
systematically investigated in \cite{MaTo08,DuMaTo08}. We briefly
recall the main results.
The stationary curve $f_\infty(w)$ satisfies the Pareto law with index $\aa$,
provided that $f_\infty$ decays like an inverse power function for large $w$,
\begin{equation}
  \label{eq.pareto}
  f_\infty(w) \propto w^{-(\aa+1)} \quad \mbox{as $w\to+\infty$} .
\end{equation}
More precisely, $f_\infty$ has Pareto index $\aa\in[1,+\infty)$ if the moments
\begin{equation}
  M_s:=\int_{\R_+} w^s \,f_\infty(w)\,dw
\end{equation}
are finite for all positive $s<\aa$, and infinite for $s>\aa$.
If all $M_s$ are finite (e.g.\ for a Gamma distribution),
then $f_\infty$ is said to possess a {\em slim tail}.

 One studies the evolution equation for the moments
\begin{equation}
  M_s(t) := \int_{\R_+} w^s\,f(w,t)\,dw ,
\end{equation}
which is obtained by integration of \eqref{bilcinetica} against $\phi(w)=w^s$,
\begin{equation}
  \label{eq.cptmomevol}
  \frac{d}{dt} M_s(t) = \frac12 \int_{\R_+ \times\R_+}\langle \phi(v^*) + \phi(w^*) \rangle f(v,t)f(w,t)\,dv\,dw - M_s(t)=: Q_+[\phi] - M_s(t)  .
\end{equation}
Using an elementary inequality for $x,y\geq0$, $s\geq1$,
\begin{equation}
  x^s+y^s \leq (x+y)^s \leq x^s+y^s + 2^{s-1}(xy^{s-1}+x^{s-1}y),
\end{equation}
one calculates for the right-hand side of (\ref{eq.cptmomevol})
\begin{equation}
  \label{eq.qest}
  \mathcal{S}(s) M_s(t) \leq Q_+[\phi] - M_s (t)\leq \mathcal{S}(s) M_s(t) + 2^{s-2}\sum_{i=1}^2\langle p_iq_i^{s-1}+p_i^{s-1}q_i\rangle M M_s^{1-1/s}(t) ,
\end{equation}
where $\mathcal{S}$ is the characteristic function
given by
\begin{equation}
\label{eq.S}
  \mathcal{S}(s) = \frac12 \Big( \sum_{i=1}^2 \langle p_i^s+q_i^s \rangle \Big) - 1.
\end{equation}
Solving (\ref{eq.cptmomevol}) with (\ref{eq.qest}),
one finds that
either $M_s(t)$ remains bounded for all times when $\mathcal{S}(s)<0$,
or it diverges like $\exp[t \mathcal{S}(s)]$ when $\mathcal{S}(s)>0$,
respectively.

The function $\mathcal{S}$ is convex in $s>0$ and $\mathcal{S}(0)=1$. It has a trivial root in $s=1$ (due to the conservation in
the mean property). It may have another non-trivial root, either in
$(1,\infty)$ or in $(0,1)$. There are three distinct cases: (i) If $s=1$ is the only root and
$\mathcal{S}(s)<0$ for all $s>1$, then all moments are
bounded, and the steady state distribution has an exponential tail; (ii) if a non-trivial root $s=\aa$ in $(1,\infty)$  exists,
moments up to the $\aa$-th moment are bounded and the steady state
distribution has a Pareto tail; (iii) if $\mathcal{S}(\aa)=0$ for some
$0<\aa<1$, then $f_\infty(w)=\delta_0(w)$, a Dirac at $w=0$.  For further details, we refer to \cite{MaTo08,DuMaTo08}, \revised{we also refer to \cite{DWB} for more complicated wealth-condensed distributions \cite{DWB}}.
%%%%

\begin{figure}
\centering
\includegraphics[width=0.5\textwidth]{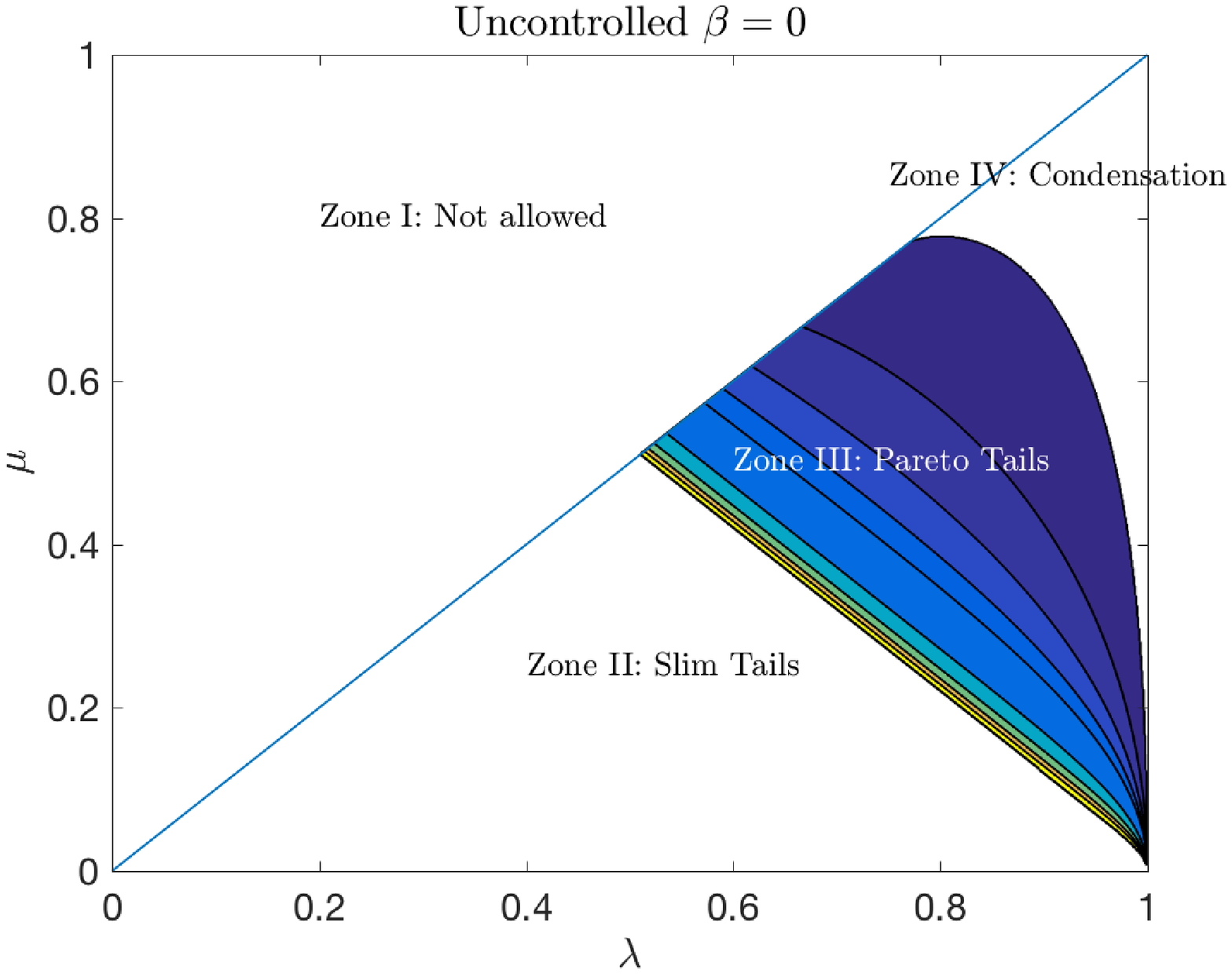}%
\includegraphics[width=0.5\textwidth]{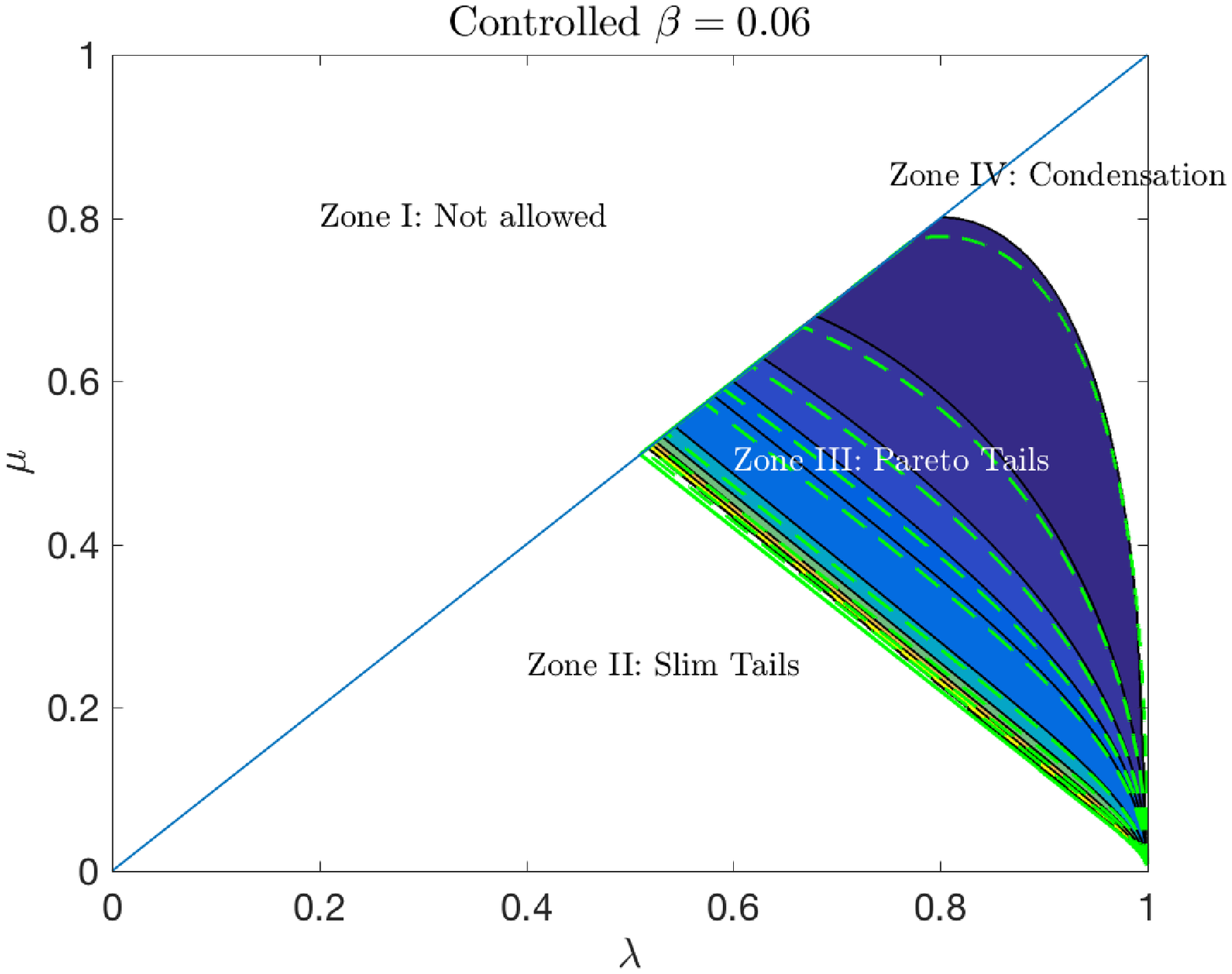}
\caption{Different tail regimes in the CPT model
  in the $\lambda$-$\mu$-plane for the uncontrolled ($\beta=0$, left) and
  controlled (\revised{quadratic cost functional}, $\beta=0.06$, right) case. The color coding in Zone III indicates the
increasing Pareto tail index, increasing from darker blue ($\aa$
close to one) to lighter
blue as $r$ increases and to yellow as $\aa\to\infty$. Dashed green curves in the right
  plot indicate the position of the contours for the uncontrolled
  case ($\beta=0$) for comparison. }
\label{fig:cptzonescontrol}
\end{figure}

We now illustrate the effect of \revised{the instantaneous control in the quadratic case} \eqref{conCPT} on the formation of the Pareto
tail. Figure~\ref{fig:cptzonescontrol} shows the
effect of the control on the formation of tails in the CPT model for
different parameters $\lambda$ and $\mu$. 

The left plot shows the
uncontrolled case. It is obtained by numerical
evaluation of the characteristic function $\mathcal{S}$.
In Zone II, $s=1$ is the only root and
$\mathcal{S}(s)<0$ for all $s>1$, hence all moments are
bounded, and the steady state distribution has an exponential tail. In
Zone III, a non-trivial root $s=\aa$ in $(1,\infty)$  exists, and
moments up to the $\aa$th moment are bounded, i.e.\ the steady state
distribution has a Pareto tail. The color coding in Zone III indicates the
increasing Pareto tail index, increasing from darker blue ($\aa$
close to one) to lighter
blue as $r$ increases and to yellow as $\aa\to\infty$. In Zone IV, there is a non-trivial root in $(0,1)$,
and condensation occurs: the steady state is a delta distribution at
zero. 

We can similarly consider the controlled case, and numerically
evaluate the characteristic function with modified mixing
parameters. The right plot in Figure \ref{fig:cptzonescontrol} shows
the effect of the control. As the control is applied
the region with slim tails (Zone II) is enlarged, while the zone with
Pareto tails (Zone III) is shifted towards the condensation zone (Zone
IV). The dashed green curves indicate the position of the contours in the
uncontrolled case for comparison.

\section{Quasi invariant limits}

\subsection{Controlled limit Fokker--Planck equation}

The analysis of \cite{MaTo08} essentially shows that the microscopic interaction \fer{eq.cpt} considered in  \cite{CPT05}  is such that the kinetic equation \fer{bilcinetica} is able to describe all interesting behaviours of wealth distribution in a
multiagent society.  

 By assuming 
 \be\label{sca1}
 \frac{1-\lambda}2 = \ve \lambda_0, \quad  \langle \eta_i^2 \rangle = \ve \sigma, \quad \tau = \ve t,
 \ee
 and a unitary average value of the initial density, it has been shown in \cite{CPT05}  that the scaled density $h(v,\tau) = f(v,t)$ satisfies in the limit $\ve \to 0$ the Fokker--Planck equation
  \be\label{FP}
 \frac{\partial h}{\partial \tau} = \frac \sigma{2}\frac{\partial^2 }
 {\partial v^2}\left( v^2 h\right) + \lambda_0 \frac{\partial }{\partial v}\left(
 (v-1) h\right).
 \ee
It is immediately recognizable that equation \fer{FP} has a unique stationary solution of unit mass, given by the
 $\Gamma$-like distribution \cite{BM,CPT05}
 \be\label{equi2}
h_\infty(v) =\frac{(\aa-1)^\aa}{\Gamma(\aa)}\frac{\exp\left(-\frac{\aa-1}{v}\right)}{v^{1+\aa}},
 \ee
  where
  $$ \aa = 1 + 2 \frac{\lambda_0}{\sigma} >1.
$$
This stationary distribution exhibits a power-law tail for large values of the wealth  variable.

The limit procedure induced by the scaling \fer{sca1}, called \emph{quasi-invariant limit} of the kinetic equation \fer{bilcinetica}, corresponds to the situation in which  are prevalent the exchanges of wealth which produce an extremely small modification  the pre-interaction wealths (\emph{grazing interactions}), but we are waiting enough time to still see the effects. 

\revised{By using the same scaling in the controlled interactions \eqref{ori2} for $\Delta t=1$, we formally obtain in the limit $\ve \to 0$ the Fokker--Planck equation
  \be\label{FPc}
 \frac{\partial h}{\partial \tau} = \frac \sigma{2}\frac{\partial^2 }
 {\partial v^2}\left( v^2 h\right) + \lambda_0 \frac{\partial }{\partial v}\left(
 (v-1) h\right)+\frac{\partial }{\partial v}\left(U[h] h\right),
 \ee
 where
\be\label{new-2}
 U[h](v,\tau)= \int_{\R_+} h(w,\tau)u_0(v,w)\,dw,
 \ee 
and $u_0(v,w)$ is the limiting value of the scaled control.} 

\revised{More precisely, by further assuming $\beta=\nu\varepsilon$, in the quadratic cost case we have
\be
u_0(v,w)=\frac{\nu}{2}(v-w),
\ee 
which gives $U[h](v,\tau)=\nu (v-1)/2$. Clearly, we obtain the same Fokker-Planck equation \eqref{FP} where now $\lambda_0$ is replaced by 
\be
\lambda_1 =\lambda_0+\nu/2. 
\label{new-2}
\ee
Since the variance of the steady state is decreasing with respect to $\lambda$,
 \[
 V(h_\infty) = \frac \sigma{2\lambda_0- \sigma},
 \]
whenever $\lambda_1 > \lambda_0$ in terms of variance the control
improves the distribution of wealth towards equality.}

\revised{
At variance, a control based on minimizing the Gini functional for $m=1$ leads to
\be
u_0(v,w) = \left\{
\begin{array}{cc}
 \displaystyle 2\nu, & \displaystyle v \geq w +\nu,  \\[+.3cm]
 \displaystyle -2\nu, & \displaystyle v \leq w -\nu,  \\[+.3cm]
 0, & {\rm otherwise}.   \\
\end{array}
\right.
\ee
In this latter case, however, the limiting equation has a different structure with respect to \eqref{FP} and we cannot compute explicitly the steady state.
}

%This can be
%easily achieved by acting on the value of the parameter $\beta$.

\subsection{Taxation-redistribution and limit Fokker--Planck equation}

The CPT model with taxation and redistribution has been proposed in \cite{BST}. There, taxation was acting on interactions \fer{eq.cpt} to take away a percentage $\delta$ of the trade wealth, to give 
\begin{equations}\label{diss}
  v{'}&=v(1-\delta) + \frac{1-\lambda}2(w-v) +   \eta_1 v,\\
w{'}&= w(1-\delta) +  \frac{1-\lambda}2(v-w)+  \eta_2 w.
 \end{equations}
Then, the wealth taken away was redistributed according to some redistribution policy, given by
a redistribution operator  of the form
 \be\label{redis}
 R_\chi^\e(f)(v,t) = \e \frac\partial{\partial v}
 \Big[ \left(\chi v - (\chi + 1)m(t)\right)f(v,t) \Big].
  \ee
Here, $m(t)$ denotes the first moment of $f$, which, in
general, makes the operator $R_\chi^\e$ nonlinear. Hence, in presence of taxation and redistribution, the weak form of the CPT-model takes the form
\begin{multline}\label{taxa}
\frac{d}{d\tau} \int_{\R_+}\phi(v) f(v,\tau) dv= \delta \int_{\R^+} \phi(v) R_\chi^\e(f)(v,t)\, dv \\
+\frac{1}{2}\left\langle \int_{\R_+ \times\R_+} f(v,\tau)f(w,\tau) \big( \phi(v{'})+\phi(w{'})-\phi(v)-\phi(w)\big)\, dv dw \right\rangle.
 \end{multline} 
Note that, by construction, the mean wealth in the system is preserved by equation \fer{taxa}.

The weight factor
multiplying the distribution function inside the square brackets in
\fer{redis} has been taken to be linear in $v$ for simplicity, also in
order to involve in the mechanism only the most meaningful moments, those
of order zero and one. Such a weight function contains only one disposable
real parameter $\chi$, a constant that characterizes the type of
redistribution, and that determines the slope of the straight line as well
as the value of $v$, whether physical or non-physical, at which the
weight itself vanishes. For $\chi > 0$ the redistribution acts in order to reduce 
inequalities proportionally to the distance from the mean wealth $m(t)$.
The other parameter has been determined by the
constraint that the redistribution operator preserves the number of agents
and actually redistributes the total amount of money that is being
collected by taxation. Further details on the redistribution operator can be found in \cite{BST, ParTosBook}. 

In a very recent paper \cite{Bisi} the quasi-invariant limit of the kinetic equation \fer{taxa} has been considered under the same scaling \fer{sca1}, by further assuming that $\delta = \kappa \ve$. The resulting Fokker--Planck equation is now
 \be\label{FPt}
 \frac{\partial h}{\partial \tau} = \frac \sigma{2}\frac{\partial^2 }
 {\partial v^2}\left( v^2 h\right) +\lambda_2 \frac{\partial }{\partial v}\left(
 (v-1) h\right),
 \ee
 \be\label{new-ll}
 \lambda_2 = \lambda_0 + \kappa(\chi +1). 
 \ee
Note that $\lambda_2 > \lambda_0$ whenever $\chi > -1$. In this case, the effect of the taxation and redistribution is to improve the distribution of wealth towards equality. 
%This can be easily achieved by acting on the values of the parameters $\kappa$ and $\chi$.

\revised{In the case of a quadratic cost functional}, apart from the different meaning of the parameters appearing in
\fer{new-2} and \fer{new-ll}, both control and taxation with
redistribution have the same effect on the quasi-invariant limit of
the CPT model, namely to increase the value of the coefficient of the drift operator in the resulting Fokker--Planck equation, thus giving a stationary distribution with smaller variance with respect to the original one. Interestingly enough, at least at the level of the Fokker--Planck equation, the effect of the taxation and redistribution (the constant $\lambda_2$) can be obtained by an instantaneous optimal control of the binary interaction simply imposing that the penalization is chosen to give $\lambda_1= \lambda_2$. This gives the identity
\be
\nu = 2 {\kappa(1+\chi)}.
\label{eq:coinc}
\ee
From this point of view, the conjecture by Piketty \cite{Piketty} is verified at the level of this simple kinetic model.

\section{A numerical comparison}
\revised{In this section we first compare the effects of the different control mechanisms induced by different cost functionals in our feedback controlled kinetic models and then analyze} the behavior of the kinetic model with local control originated by a quadratic cost functional with the kinetic model based on a global redistribution mechanism. All models have been solved using a direct Monte Carlo simulation method (see \cite{ParTosBook} for more details). The noise term has been taken as $\eta_i=\pm \mu$, where each sign comes with probability $1/2$. Therefore, we have $\langle \eta_i^2 \rangle =\mu^2$ in \eqref{sca1}. The number of simulated sample agents has been fixed to $N=5\times 10^4$ and standard averaging procedures have been used after the steady state has been reached to reduce the statistical fluctuations.

\begin{figure}[t]
\centering
\includegraphics[width=0.5\textwidth]{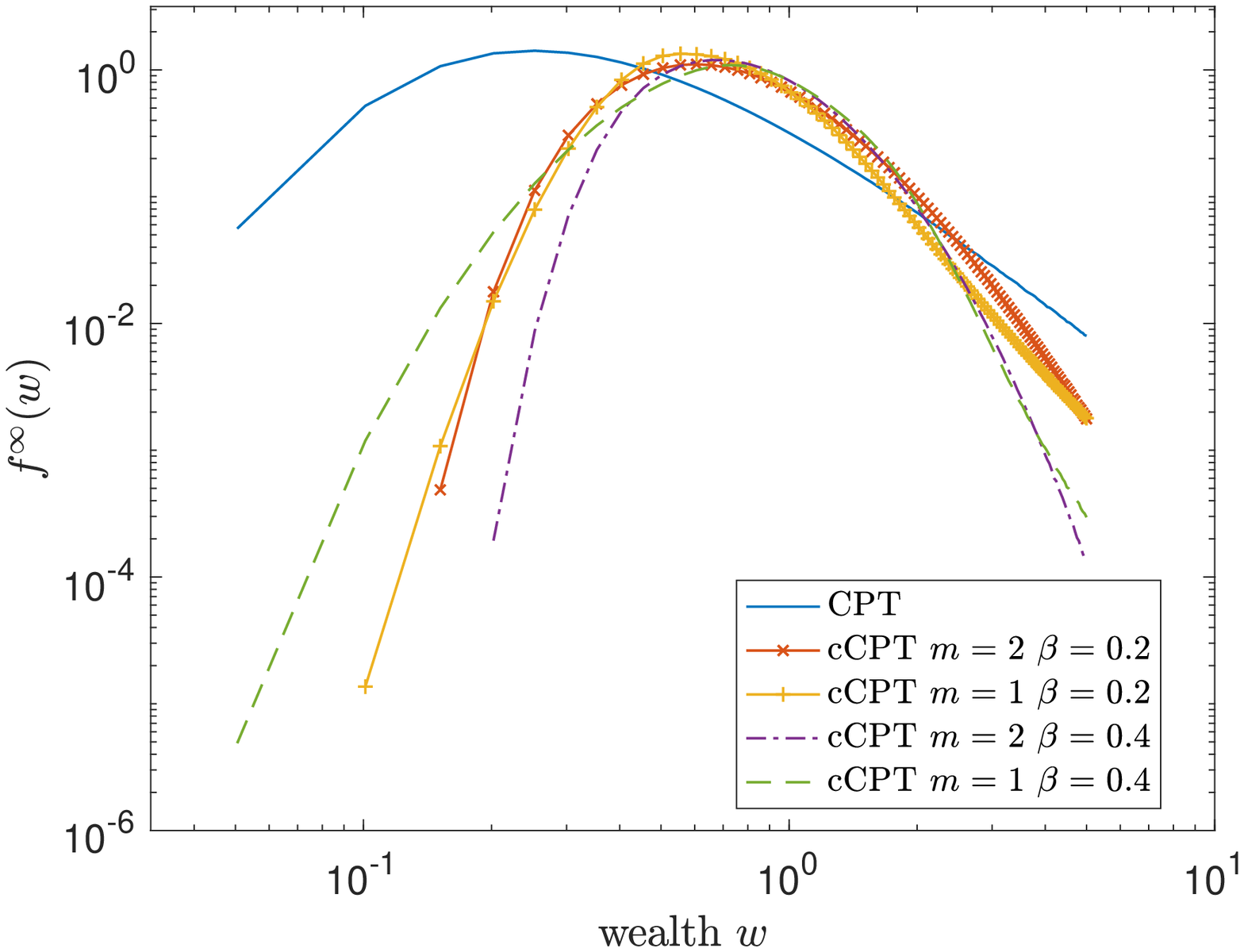}\,\,%
\includegraphics[width=0.48\textwidth]{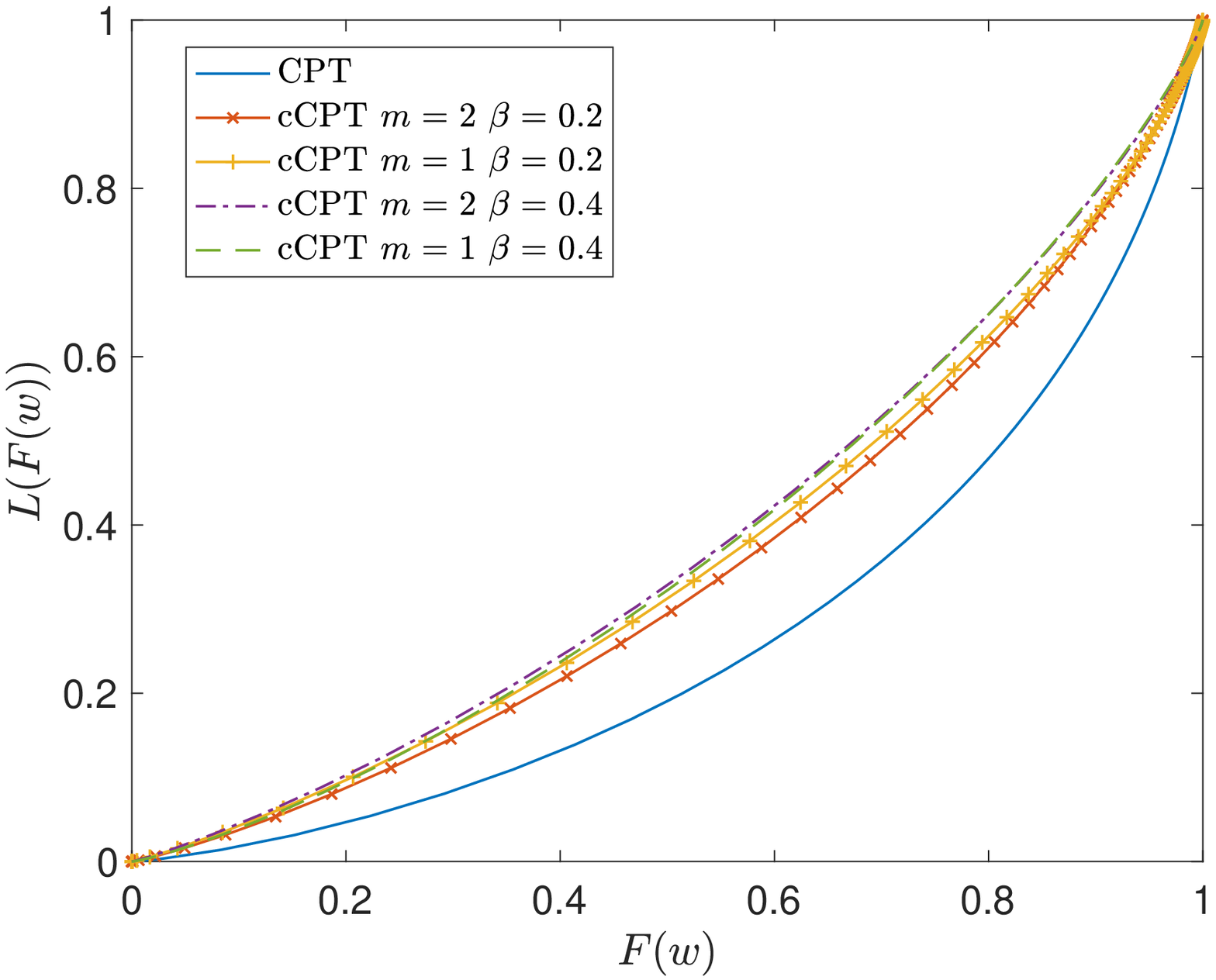}
\caption{Test 1. Behavior of the controlled (cCPT) model in the case $m=1$ and $m=2$ for various values of $\beta$. Here $\mu=0.25$ and $\lambda=0.95$. Asymptotic behavior of the solution in loglog scale (left) and  corresponding Lorentz curves (right).}
\label{fig:test0}
\end{figure}

\revised{\subsection{Test 1. The effects of different feedback controls}
First we compare the effects of the different control induced by the choice of the cost functional in \eqref{eq:cost}. More precisely we compare the controlled kinetic model (cCPT) defined by \eqref{ori2} where the feedback control is defined by \eqref{conCPT} for $m=2$ and by \eqref{conCPT2} for $m=1$. We fix the strength of noise $\mu=0.25$ and select $\lambda=0.95$.  With these choices we are in the power law asymptotic region of the CPT model (see Figure \ref{fig:cptzones}). The maximum admissible control value for $\beta$ is about $0.47$ for $m=2$. Initially each sample agent has a wealth $w=1$ so that $f_0(w)=\delta(1)$. The results are reported in Figure \ref{fig:test0} for $\beta=0.2$ and $\beta=0.4$. Both controls mechanisms provide a marked reduction of inequalities in the system, in particular the reduction of the Pareto index in the power law tail is proportional to the penalization term $\beta$ and comparable in the two models (see Figure \ref{fig:test0}, left). On the other hand, the effects of the different controls processes are clearly evident for lower values of the wealth. Increasing $\beta$ for $m=1$ implies a taxation/redistribution process for larger differences in wealth (accordingly to Figure \ref{fig:control}) which as a results gives less opportunities for agents with low wealth values to benefit of the inequality reduction process.} 

\revised{Next, to emphasize the reduction of wealth inequality in the same Figure (right) we have also plotted the Lorentz curve defined as
\[
L(F(w)) = \int_0^w f^{\infty}(v) v\,dv,\quad F(w) = \int_0^w f^{\infty}(v)\,dv, 
\] 
since we have $\int_0^\infty f^{\infty}(v) v\,dv=1$.
The Gini coefficient $G_1$ in \eqref{eq:Gini} can then be thought of as the ratio of the area that lies between the line of equality (the line $y=x$ of perfect equality) and the Lorenz curve over the total area under the line of equality. In our test case we have a value of $G_1\approx 0.46$ in the uncontrolled case. For $\beta=0.2$ it is quite evident that the feedback control with $m=1$ yields a stronger reduction of the Gini coefficient ($G_1\approx 0.3$ for $m=2$ and $G_1\approx 0.28$ for $m=1$), whereas for $\beta=0.4$ the two control gives analogous results ($G_1\approx 0.25$ for both models).} 

\subsection{Test 2. Local control and global redistribution}
\revised{Next we compare the kinetic model obtained by minimization of a quadratic cost functional defined by \eqref{CPTc}  with the corresponding model based on a global taxation/redistribution process in \eqref{diss}-\eqref{redis}. The simulation is performed in the quasi-invariant scaling defined by \eqref{sca1} together with the further scaling $\beta=\nu\varepsilon$ and $\delta = \kappa\varepsilon$.} In all test cases, the parameters in the two models are related by assumption \eqref{eq:coinc} so that in the limit $\varepsilon\to 0$ their solution should coincide.
We report the results obtained with the different models for various values of the scaling parameter $\varepsilon$. In this way for $\varepsilon = \mathcal{O}(1)$ we can emphasize the different behavior of the local control when compared to a global redistribution policy, whereas for $\varepsilon \ll 1$ we can verify the asymptotic procedure that lead to the same Fokker-Planck equation. 

\begin{figure}[t]
\centering
\includegraphics[width=0.5\textwidth]{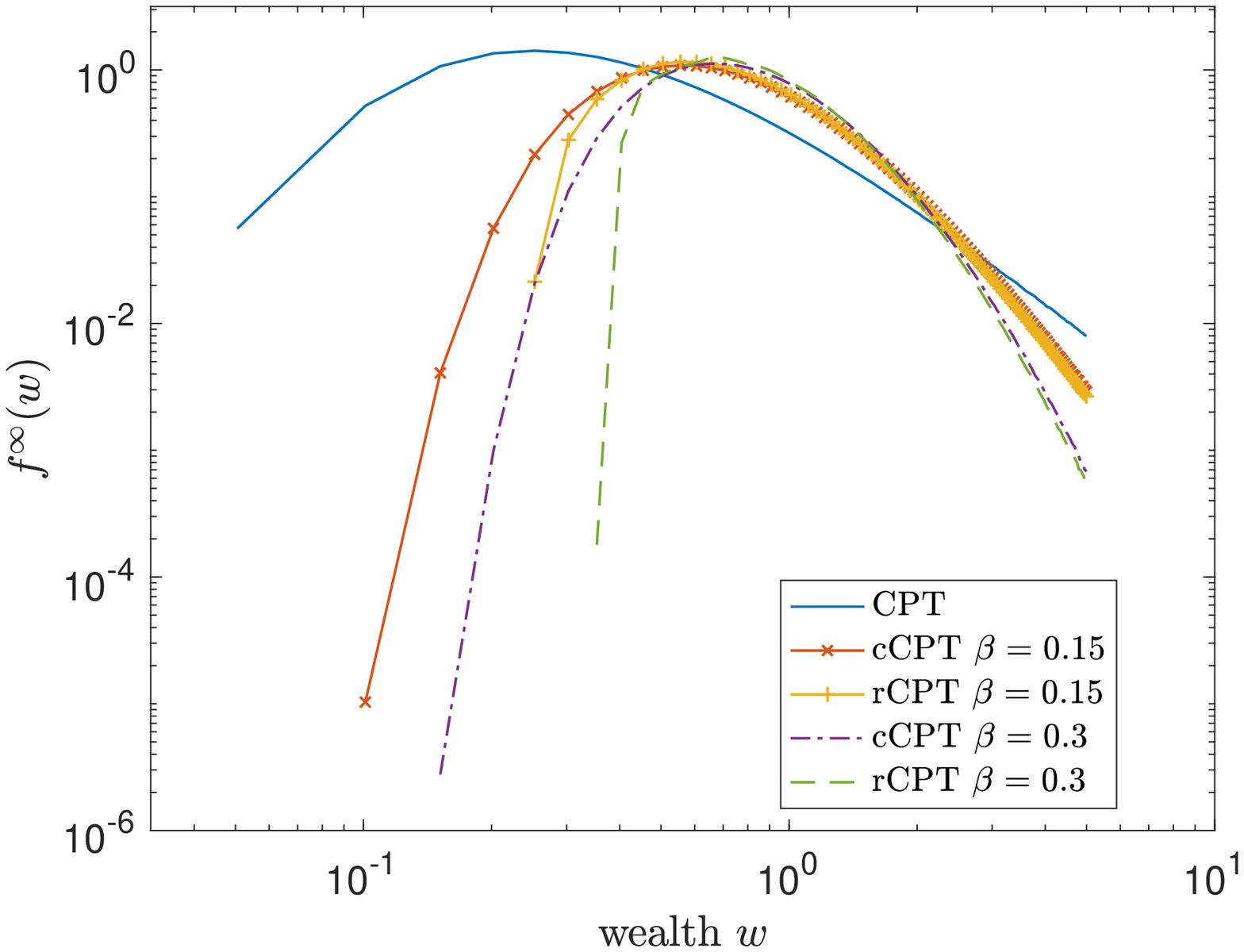}\,\,%
\includegraphics[width=0.48\textwidth]{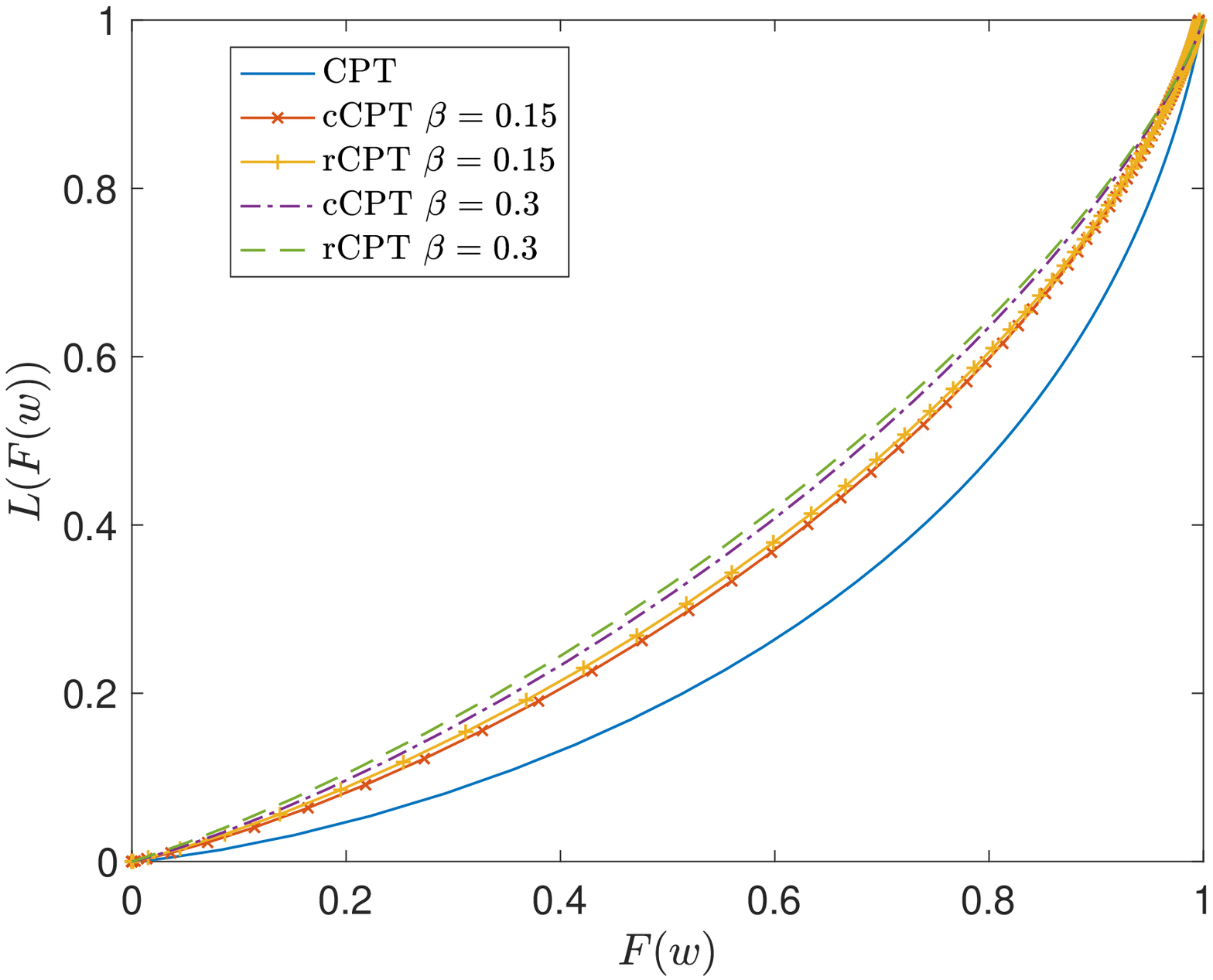}
\caption{Test 2. Behavior of the controlled (cCPT) for $m=2$ and redistributed (rCPT) models for various values of $\beta$. Here $\mu=0.25$, $\lambda=0.95$, $\delta=\lambda\beta/2$, $\chi=0$. Asymptotic behavior of the solution in loglog scale (left) and corresponding Lorentz curves (right).}
\label{fig:test1}
\end{figure}

\paragraph{The $\varepsilon=\mathcal{O}(1)$ regime.}
In the first test case we compare the controlled (cCPT) model for $m=2$ and the redistributed (rCPT) model in absence of scaling, or equivalently taking $\varepsilon=1$. We set $\mu=0.25$ and $\lambda=0.95$ as in Test 1 and the same initial data.  For the redistributed model we fix $\delta = \lambda \beta /2$, so that the taxation process of the two models is the same in each binary interaction, and choose $\chi=0$ so that the redistribution process is independent from the wealth. 

As expected, with these choices the two models show a rather similar behavior. 
The results of the corresponding stationary solutions are reported in Figure \ref{fig:test1} (left) for $\beta=0,0.15$ and $0.3$. The different slopes of the tails clearly show how both models are capable to reduce the inequalities in the wealth distribution. Note that the models behavior is different for small values of the wealth, since in the redistributed model the density of agents with wealth below $\delta$ is exactly equal to zero. In Figure \ref{fig:test1} (right) we report the corresponding Lorentz curves. 

%The two models yields analogous results for this cho
%In our test case we have $G_1\approx 0.32$ and $G_1\approx 0.27$ in the controlled cases for $\beta=0.15$ and $\beta=0.30$, respectively. 

\begin{figure}[t]
\centering
\includegraphics[width=0.5\textwidth]{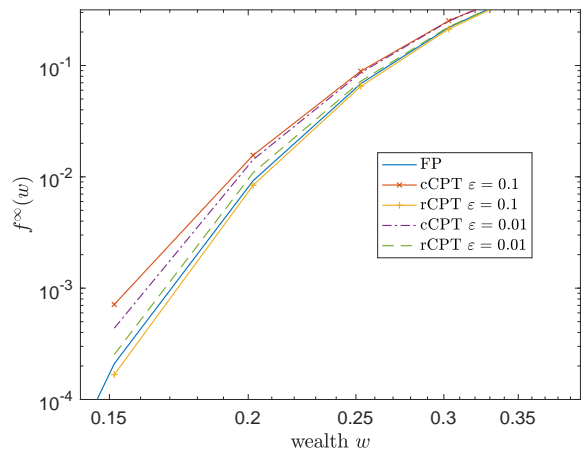}%
\caption{Test 2. Scaling limit of the controlled (cCPT) for $m=2$ and redistributed (rCPT) models for various values of $\varepsilon$. Here $\mu=0.25$, $\lambda=0.95$, $\beta=0.15$, $\delta=\lambda\beta/2$, $\chi=0$. Asymptotic behavior of the solution in loglog scale for small values of the wealth.}
\label{fig:test2}
\end{figure}

\paragraph{The limit $\varepsilon\to 0$.}
Finally, we consider the scaling process that leads from the kinetic Boltzmann models to their corresponding Fokker-Planck descriptions \eqref{FPc} and \eqref{FPt}. We consider the same data as before but for a fixed value of $\beta = 0.15$ and various values of the scaling parameter $\varepsilon=0.1$ and $0.01$. For this choice of parameters in the limit $\varepsilon\to 0$ we obtain a Pareto index $r=1.8$ in the uncontrolled case and $r=4.2$ in the controlled case with $\beta=0.15$.

The results are reported in Figure \ref{fig:test2}. Since for large values of the wealth the two models give very similar results and show the same power law behavior of the limit Fokker-Planck model,
to remark the differences we considered a region of the density function close to the left boundary $w=0$. The convergence of the models towards the analytic steady state of the Fokker-Planck model is evident.

\section{Conclusions}
In this paper, we introduced a possible alternative to the standard
taxation and redistribution rules, which relies in a suitable control
applied to the microscopic trades describing the wealth distribution
of a multi-agent system. The constrained system is then approximated
by a finite time horizon strategy which allows to embed explicitly the
control in the interaction rules. \revised{We emphasize that the resulting form 
of the control is closely related to the choice of the cost functional.
Different cost functionals originate different taxation/redistribution
strategies. We analyze in details the case of a cost functionals which aims
at minimizing the variance of the wealth distribution and the case 
of a cost functional which minimizes the well-known Gini indicator.}  
The corresponding kinetic models
based on binary interactions can then be derived and show that the
control is able to modify the corresponding Pareto tails. This can be
further analyzed with the aid of some numerical simulations by considering 
the corresponding quasi-invariant Fokker-Planck limit and its relationship 
with previous models based on global taxation and redistribution.      

\section*{Acknowledgements}
B.~D{\"u}ring acknowledges support by the Leverhulme Trust research
project grant {\em Novel discretisations for higher-order nonlinear PDE} (RPG-2015-69).
L.~Pareschi acknowledges the support of the MIUR-DAAD program \emph{Mean field games for socio economic problems}. G.~Toscani acknowledges the support of the Italian Ministry of Education, University and Research (MIUR): \emph{Dipartimenti di Eccellenza Program (2018--2022) - Dept. of Mathematics ``F. Casorati'', University of Pavia}.


\begin{thebibliography}{99}

\bibitem{AHP} Albi, G.,  Herty, M.\ and Pareschi, L.  Kinetic description of optimal control problems and applications to opinion consensus. \emph{Commun.\ Math.\ Sci.}, \textbf{13} (6) 1407–-1429, (2015).

\bibitem{APZ} Albi, G.,  Pareschi, L.\ and Zanella, M. Boltzmann-type control of opinion consensus through leaders. \emph{ Phil.\ Trans.\ R.\ Soc.\ A} \textbf{372}, 20140138, (2014).

\bibitem{AlbiEtAl16} Albi, G., Pareschi, L., Toscani, G.\ and Zanella,
  M. Recent advances in opinion modeling: control and social
  influence.  In  {\em Active Particles, Volume 1: Theory, Models, Applications}. N.\ Bellomo, P.\ Degond, E.\ Tadmor, Eds. Ch.2, pp. 49--98.  Birkh\"auser, Boston, (2017).

\bibitem{Angle} Angle, J. The surplus theory of social stratification
  and the size distribution of personal wealth. {\em Social Forces}, {\bf
    65}, 293--326, (1986). 
    
\bibitem{Bisi}
Bisi, M. Some kinetic models for a market economy. \emph{Boll.\ Unione Mat.\ Ital.} \textbf{10}, 143–-158, (2017).    
    
\bibitem{BST} 
Bisi, M., Spiga, G.\ and Toscani, G. Kinetic models of conservative economies with wealth redistribution.
\emph{Commun.\ Math.\ Sci.} \textbf{7} (4) 901--916, (2009).   

\bibitem{BM}
Bouchaud, J.F.\ and M\'ezard, M. Wealth condensation in a
simple model of economy. \emph{Physica A\/}, {\bf 282},  536--545, (2000). 


\bibitem{CaBo:04}
{Camacho, E.\ and Bordons, C.}, {\em Model predictive control}, Springer, USA,
  2004.
  
\bibitem{Cerc88} Cercignani, C. 
{\em The Boltzmann Equation and its Applications}. Springer Series in
Applied Mathematical Sciences, Vol.\ 67, New York, (1988).

\bibitem{CIP94} Cercignani, C., Illner, R.\ and Pulvirenti, M. 
{\em The Mathematical Theory of Dilute Gases}, Springer Series in
Applied Mathematical Sciences, Vol.\ 106, New York, (1994).

\bibitem{Cha} 
Chakraborti, A. Distributions of money in models of market economy. \emph{Int.\ J.\ Modern Phys.\ C} \textbf{13},  1315--1321, (2002).

\bibitem{CC} Chakraborti, A.\ and  Chakrabarti, B.K. Statistical Mechanics of Money: Effects of Saving Propensity. \emph{Eur.\ Phys.\ J.\ B} \textbf{17}, 167--170, (2000).

\bibitem{CPT05} Cordier, S., Pareschi, L.\ and Toscani, G. On a kinetic
  model for a simple market economy, {\em J.\ Statist.\ Phys.}, \textbf{120}, (2005)

%\bibitem{DJR}  Degond, Pierre ;  Liu, Jian-Guo ;  Ringhofer, Christian . Evolution of wealth in a non-conservative economy driven by local Nash equilibria.
% Philos. Trans. R. Soc. Lond. Ser. A Math. Phys. Eng. Sci.  372  (2014),  no. 2028, 20130394, 15 pp.
        
\bibitem{DJR} Degond, P.,  Liu, J-G.,  Ringhofer, C. Evolution of the distribution of wealth in an economic environment driven by local Nash equilibria.
 \emph{J. Stat. Phys.},  {\bf 154}, 751--780 (2014).
        

\bibitem{DWB} Devitt-Lee A., Wang H., Li J., and Boghosian B.  A Nonstandard Description of Wealth Concentration in Large-Scale Economies, \emph{SIAM J. Appl. Math.}, 78(2), 996--1008, (2017).

\bibitem{DuMaTo08} D{\"u}ring, B., Matthes,
  D.\ and Toscani, G. Kinetic equations modelling wealth
  redistribution: a comparison of approaches, {\em Phys.\ Rev.\ E\/}
  \textbf{78}(5), 056103, (2008).

\bibitem{DMT09} D{\"u}ring, B., Matthes, D.\ and Toscani, G. 
A Boltzmann-type approach to the formation of wealth distribution curves.
{\em Riv.\ Mat.\ Univ.\ Parma (8)} \textbf{1}, 199-261, (2009).

\bibitem{DuTo07} D{\"u}ring, B.\ and Toscani, G. 
Hydrodynamics from kinetic models of conservative economies.
{\em Physica~A} \textbf{384}(2), 493-506, (2007).

\bibitem{HSP}
Herty, M., Steffensen, S. \ and Pareschi, L. Mean--field control and Riccati equations, \textit{Networks and Heterogeneous Media}, \textbf{10}: 699--715,  (2015). 

\bibitem{krstic1995}
{Krstic, M., Kanellakopoulos, I. \ and Kokotovic, P.V.}, {\em Nonlinear and
  adaptive control design}, John Wiley and Sons Inc., New York, 1995.

\bibitem{MaTo08} Matthes, D.\ and Toscani, G. 
On steady distributions of kinetic models of conservative economies.
{\em J.\ Stat.\ Phys.} \textbf{130}(6), 1087-1117, (2008). 

\bibitem{ParTosBook} Pareschi,  L.\ and Toscani, G.  
{\em Interacting multiagent systems: kinetic equations and Monte Carlo
methods}, Oxford University Press, Oxford, UK, (2014).

\bibitem{PT-kno}
 Pareschi, L.\ and  Toscani, G. Wealth distribution and collective knowledge. A Boltzmann approach,  \emph{Phil.\ Trans.\ R.\ Soc.\ A}  \textbf{372}, 20130396, (2017).

%\bibitem{PZ-str}
% Pareschi, L. and  Zanella, M. Structure Preserving Schemes for Nonlinear Fokker--Planck Equations and Applications,  %\emph{J. Sci. Comp}, 1--26, (2017).
 

\bibitem{Par} Pareto, V. {\em Cours d'\'economie politique}, Tome II
  Livre III. F. Pichon, Imprimeur-\'Editeur, Paris, France, (1897).
  
\bibitem{PR} 
Perversi, E.\ and Regazzini, E. Inequality and risk aversion in economies open to altruistic attitudes.
\emph{Math.\ Models Methods Appl.\ Sci.}  \textbf{26}, 1735--1760, (2016).   

\bibitem{Piketty} Piketty, T. {\em Capital in the Twenty-First Century},
  Harvard University Press, Cambridge MA, (2013).

\bibitem{Donadio} Scalas, E., Garibaldi, U.\ and Donadio, S. Statistical
  equilibrium in simple exchange games I: Methods of solution and
  applications to the Bennati-Dr\v{a}gulescu-Yakovenko (BDY) game. {\em Eur.\ Phys.\ J.\ B}, {\bf 53}, 267--272 (2006). 

\bibitem{Sontag1998aa}
{ Sontag, \sc E.~D.}, {\em Mathematical control theory}, vol.~6 of Texts in
  Applied Mathematics, Springer-Verlag, New York, second~ed., 1998.

\bibitem{To09}
Toscani, G. Wealth redistribution in conservative linear kinetic models with taxation. \emph{Europhys.\ Letters} \textbf{88}, (1)  10007, (2009).

\bibitem{To17}
Toscani, G. Continuum models in wealth distribution. \emph{Rend.\ Lincei Mat.\ Appl.} \textbf{28} 451–-461 (2017)     

\bibitem{Yakovenko} Dr\v{a}gulescu, A.\ and Yakovenko, V., Statistical
  mechanics of money. {\em Eur.\ Phys.\ J.\ B}, {\bf 17}, 723--729
  (2000).
 \end{thebibliography}
\end{document}